\documentclass[12pt,draftcls,onecolumn]{IEEEtran}

\newtheorem{TT}{Theorem}

\ifCLASSINFOpdf
\else
\fi

\ifCLASSOPTIONcompsoc
\usepackage[nocompress]{cite}
\else
\usepackage{cite}
\fi

\usepackage{amsmath}
\usepackage{amssymb}
\usepackage{graphicx}
\usepackage{epstopdf}
\usepackage{mathtools}
\usepackage{xcolor}
\begin{document}

\title{Discrete Spectrum Analysis of Vector OFDM Signals}

%\author{\IEEEauthorblockN{Xin Meng, Xiang-Gen Xia, and Xiqi Gao}}

\author{Xiang-Gen Xia, \IEEEmembership{Fellow}, \IEEEmembership{IEEE} 
       and Wei Wang, \IEEEmembership{Senior Member}, \IEEEmembership{IEEE}

\thanks{X.-G. Xia is with the Department of Electrical and Computer Engineering,
  University of Delaware, Newark, DE 19716, USA (e-mail: xxia@ece.udel.edu).
  W. Wang is with Peng Cheng Laboratory,
  Shenzhen, China (e-mail: wei\_wang@ieee.org). W. Wang is the correspondence
author.}

%\author{\IEEEauthorblockN{Xin Meng, Xiang-Gen Xia, and Xiqi Gao}\\
%\IEEEauthorblockA{
%$^{*}$National Mobile Communications Research Laboratory, Southeast University, Nanjing, 210096, China \\
%$^{\dag}$Institute of Electronics, Xidian University, Xi'an, Shaanxi 710071, China \\
%$^{\ddag}$Department of Electrical and Computer Engineering, University of Delaware, Newark, DE 19716, USA
%}
}

\date{}

\maketitle

%\vspace*{-33pt}

\begin{abstract}
  Vector OFDM (VOFDM) is equivalent to  OTFS and is good
 for time-varying channels. However, due to its vector form, its signal spectrum
 is not as clear as that of the conventional OFDM.
 In this paper, we
  study the discrete spectrum of discrete VOFDM  signals. We obtain
  a linear relationship between a vector of information symbols
  and a  vector of the same size
  of components evenly distributed in the discrete VOFDM signal spectrum, and 
  show that
  if a vector of information symbols is set to $0$,
  then a corresponding vector of the same size of
  the discrete VOFDM signal spectrum is $0$ as well,
  where the components of the $0$ vector are not together
  but evenly distributed 
  in the spectrum. With the linear relationship, the
  information symbol vectors can be locally
  precoded so that any of the discrete
  spectrum of VOFDM signals can be set to $0$,  similar to  that of
  the conventional OFDM signals.
  These results are verified by simulations. 
\end{abstract}


%\newpage

\section{Introduction}\label{sec1}
Vector OFDM (VOFDM) was first proposed in 2000 \cite{xia1, xia2, xiabook} and 
has had more investigations in,
for example, \cite{han, cheng, xia5}. There  have been  its equivalent
systems, such as OSDM \cite{ODSM} and A-OFDM \cite{AOFDM}. 
VOFDM converts an intersymbol interference (ISI) channel to
multiple vector subchannels where there is no ISI across vector
subchannels and in each vector subchannel, the ISI is limited
by the vector size. VOFDM is in the middle of OFDM and single carrier 
frequency domain equalizer (SC-FDE) \cite{xia5} and is the most general modulation
dealing with ISI.
Furthermore, it was shown in \cite{xiabook, xia5, xia3, xia6} that VOFDM is good for
time-varying channels due to its vector-wise demodulation. 
Interestingly,  it has been recently shown in, for example, 
 \cite{otfs2, otfs3, xia3, osdm}, that VOFDM is also equivalent
to OTFS \cite{otfs1, otfs4} for delay Doppler channels,
which has been an active research topic lately. For delay Doppler
channels, a recent study is \cite{xia7}. 

It is known that the discrete spectrum of a conventional OFDM signal
is clear and the information symbols located at the subcarriers
are the responses at the corresponding discrete frequencies. Thus, it is
easy to avoid certain frequrencies by setting the information symbols $0$ at these
frequencies  and therefore is flexible in the design
in practice. Due to the vector form in VOFDM, 
the spectrum of a VOFDM signal is not as clear as that of an OFDM signal.
In this paper, we
study the discrete spectrum of discrete VOFDM  signals.
We first obtain
  a linear relationship between a vector of information symbols
  and a  vector of the same size
  of components evenly distributed in the
  discrete VOFDM signal spectrum.
  The linear relationship is in fact the $M$-point DFT
  of a modulated information symbol vector, where $M$ is the
  vector size. 
  With this linear relationship, it is easy
  to see  that
  if a vector of information symbols is set to $0$,
  then a corresponding vector of the same size of
  the discrete VOFDM signal spectrum is $0$ as well,
  where the components of the $0$ vector are not together
  but evenly distributed 
  in the spectrum. Also with the linear relationship, the
  information symbol vectors can be locally
  precoded so that any of the discrete
  spectrum of VOFDM signals can be set to $0$,  similar to  that of
  the conventional OFDM signals. Thus, it is also flexible
  in a practical design to avoid certain frequencies or frequency bands.
  Since the precoding is done for a vector (or vector-wisely), when the vector size is not
  large, the precoding complexity is low. 

  This paper is organized as follows. In Section \ref{sec2}, we
  derive the discrete spectrum of a VOFDM signal and present the 
  precoding to pre-satisfy a $0$ spectrum property of VOFDM signals.
  In Section \ref{sec3}, we present some simple
  simulation results to verify the theoretical results.
  In Section \ref{sec3}, we conclude this paper. 
  
  \section{Discrete Spectrum and Precoding}\label{sec2}
  We first formulate a discrete VOFDM signal \cite{xia1, xia2, xia5, xia6}.
  Let $M$ be the vector size and $N$ be the IFFT size.
  Let
\begin{equation}\label{0}
  {\bf x}_k=(x_k(0),x_k(1),\cdots, x_k(M-1))^T, \,\,0\leq k\leq N-1,
  \end{equation}
  be information symbol vectors to be transmitted, where $^T$ stands for the
  transpose. For the $m$th components of these $N$ vectors, $0\leq m\leq M-1$, 
  along the index $k$ direction, the $N$-point IFFT is
$$
  \mbox{IFFT}_N (x_0(m), x_1(m), \cdots, x_{N-1}(m))
$$
    \begin{equation}\label{1}
    \stackrel{\Delta}{=} (X_0(m),X_1(m),\cdots, X_{N-1}(m)).
  \end{equation}
  Let
\begin{equation}\label{2}
  {\bf X}_n = (X_n(0), X_n(1), \cdots, X_n(M-1))^T,\,\, 0\leq n\leq N-1,
\end{equation}
and
$$
  {\bf X}=(X(0),X(1),\cdots, X(MN-1))^T
$$
 \begin{equation}\label{3}
  =({\bf X}_0^T, {\bf X}_1^T, \cdots, {\bf X}_{N-1}^T)^T.
  \end{equation}
Then, ${\bf X}$ is a discrete VOFDM signal
before the cyclic prefix (CP) insertion.
  Since this paper is only interested in the spectrum of a VOFDM signal,
  for simplicity, we do not consider a CP for a VOFDM signal.

  We now calculate the discrete spectrum of the discrete VOFDM signal ${\bf X}$
  by taking its $MN$-point FFT:
  $${\bf y}=(y(0), y(1), \cdots, y(MN-1))^T=\mbox{FFT}_{MN}({\bf X})$$
$$
    \stackrel{\Delta}{=}
    (y_0(0), y_1(0), \cdots, y_{N-1}(0), y_0(1), y_1(1), \cdots,
        y_{N-1}(1),
    $$
\begin{equation}\label{4}
    \cdots, y_0(M-1), y_1(M-1),\cdots, y_{N-1}(M-1))^T.
  \end{equation}
  Finally, let
  \begin{equation}\label{5}
    {\bf y}_k=(y_k(0),y_k(1),\cdots,y_k(M-1))^T,\,\,0\leq k\leq N-1,
    \end{equation}
  which are the spectrum vectors of size $M$ and their
  components are evenly distributed (not consecutively located)
  in the discrete spectrum ${\bf y}$ of the VOFDM signal ${\bf X}$.
  
  As one can see that the above discrete spectrum vectors ${\bf y}_k$ are
  not obviously
  related to the original information vectors ${\bf x}_k$ as that for
  the OFDM signals, i.e., when $M=1$ and ${\bf y}_k={\bf x}_k$ for $0\leq k\leq N-1$.
  We next find a relationship between vectors ${\bf y}_k$ and ${\bf x}_k$ for $0\leq k\leq N-1$ by writing out ${\bf y}$ in (\ref{4}) as follows.
  First, let $W_K= e^{-2\pi j/K}$ for a positive integer $K$. Then,
  for $0\leq k_1\leq M-1$ and $0\leq k_2 \leq N-1$, 
  \begin{eqnarray*}
  & &    y(k_1 N+k_2)\\
  & = &  \frac{1}{\sqrt{MN}}  \sum_{n=0}^{MN-1} X(n) W_{MN}^{n(k_1N+k_2)} \\
  &  = &   \frac{1}{\sqrt{MN}}\sum_{n_1=0}^{M-1} \sum_{n_2=0}^{N-1} X_{n_2}(n_1)
                    W_{MN}^{(n_2M+n_1)(k_1N+k_2)}\\
       &  = &   \frac{1}{\sqrt{MN}} \sum_{n_1=0}^{M-1} \sum_{n_2=0}^{N-1} X_{n_2}(n_1)
          W_{MN}^{n_1k_2}W_M^{n_1k_1}W_N^{n_2k_2}\\
          & = &    \frac{1}{\sqrt{M}} \sum_{n_1=0}^{M-1}
          \left(  \frac{1}{\sqrt{N}}\sum_{n_2=0}^{N-1} X_{n_2}(n_1)
        W_N^{n_2k_2}\right) W_M^{n_1k_1}W_{MN}^{n_1k_2}\\
        & \stackrel{(a)}{=}&  \frac{1}{\sqrt{M}}\sum_{n_1=0}^{M-1} x_{k_2}(n_1) W_{MN}^{n_1(k_1N+k_2)},
  \end{eqnarray*}
   where Step (a) is from (\ref{1}).  This leads to
  the following relationship between ${\bf y}_k$ and ${\bf x}_k$:
  \begin{equation}\label{6}
    y_{k}(m)=\frac{1}{\sqrt{M}}\sum_{n=0}^{M-1} \left(x_k(n)W_{MN}^{nk}\right) W_{M}^{nm},
  \end{equation}
for     $0\leq m\leq M-1, 0\leq k\leq N-1$.
  The above linear relationship is in fact the $M$-point DFT of
  the information symbols $x_k(n)$ modulated
  by $W_{MN}^{nk}$ in terms of $n$, $0\leq n\leq M-1$.
  Interestingly this is related to the relationship (5.2)
  in \cite{xia2} when the channel is ideal and does not
  have any additive noise. 
  From (\ref{6}), we immediately have the following theorem.

  \begin{TT}
    The $k$th discrete spectrum vector ${\bf y}_k$ in (\ref{5})
    of a discrete VOFDM signal
    and the $k$th information symbol vector  ${\bf x}_k$ in (\ref{0})
    satisfy the
    linear relationship (\ref{6}), where $0\leq k\leq N-1$. If 
    vector ${\bf x}_{k_0}=0$ for some $0\leq k_0\leq N-1$, then
    vector ${\bf y}_{k_0}=0$.
  \end{TT}
  
  One can easily see from (\ref{6}) that for the conventional OFDM,
  i.e., when $M=1$, we have ${\bf y}_k={\bf x}_k$ for $0\leq k\leq N-1$
  as we have mentioned earlier. 
  However, when $M>1$, VOFDM is different from
  OFDM and if any component of information symbol vector ${\bf x}_k$
  is not $0$, 
  then none of the spectrum components in ${\bf y}_k$
  vanishes,  i.e., ${\bf y}_k$ is full,
  for any $0\leq k\leq N-1$. 
%  \section{Precoding for Achieving Zero Frequency Components} 
  Although this is the case, 
  if some of the frequencies are to avoid for the system, i.e.,
  some components in ${\bf y}_k$ are set to $0$, then one may precode
  the information symbol vector ${\bf x}_k$ so that (\ref{6}) holds, i.e.,
  the pre-set frequency components in ${\bf y}_k$ are indeed $0$. 
  This can be obtained as follows.

  Consider the last a few components of the spectrum vector ${\bf y}_k$
  are set to $0$ for some $0\leq k\leq N-1$, i.e.,
  \begin{equation}\label{7}
    {\bf y}_k=({\bf y}_{k,1}^T, {\bf y}_{k,2}^T)^T= ({\bf y}_{k,1}^T, 0,\cdots,0)^T,
  \end{equation}
  where ${\bf y}_{k,2}=(0,\cdots,0)^T$ or simply denoted  by 
  ${\bf y}_{k,2}=0$. Let the information symbol vector
  \begin{equation}\label{8}
    {\bf x}_k=({\bf x}_{k,1}^T, {\bf x}_{k,2}^T)^T,
  \end{equation}
  where ${\bf x}_{k,2}$ has the same size as ${\bf y}_{k,2}$
  and is to be determined below,  and ${\bf x}_{k,1}$
  is the information symbol vector that contains information symbols to
  transmit.

  The {\em precoding} is that for any given information symbol vector
  ${\bf x}_{k,1}$, we want to determine vector ${\bf x}_{k,2}$ such that
  (\ref{6}) holds for ${\bf y}_k$ in (\ref{7}) and ${\bf x}_k$ in (\ref{8}).
  And ${\bf x}_{k,2}$ is called a precoded signal. We next show how this
  precoding is done. 
  
  Let
  $$
  {\bf W}_M=\frac{1}{\sqrt{M}}\left(W_M^{mn}\right)_{0\leq m,n\leq M-1}
  $$
  be the $M$-point DFT matrix, ${\bf I}_M$ be the $M\times M$ identity matrix,
  and
  \begin{equation}\label{9}
    {\bf W}_M=\left( {\bf W}_{M,1}, {\bf W}_{M,2}\right) \mbox{ and }
    {\bf I}_M=\left( {\bf I}_{M,1}, {\bf I}_{M,2}\right),
  \end{equation}
  where the numbers of the columns of submatrices ${\bf W}_{M,2}$
  and ${\bf I}_{M,2}$ are the same as  the dimension of
  vector ${\bf y}_{k,2}$. Then, (\ref{6}) is equivalent to
  %\begin{eqnarray}
  $$
    \left( {\bf I}_{M,1}, -{\bf W}_{M,2}\right)
    \left( \begin{array}{c} {\bf y}_{k,1}\\ \hat{\bf x}_{k,2} \end{array}\right)
    =   \left( {\bf W}_{M,1}, -{\bf I}_{M,2}\right)
    \left( \begin{array}{c} \hat{\bf x}_{k,1}\\ {\bf y}_{k,2} \end{array}\right)
    $$
   \begin{equation} \label{10}
    =   \left( {\bf W}_{M,1}, -{\bf I}_{M,2}\right)
    \left( \begin{array}{c} \hat{\bf x}_{k,1}\\ 0 \end{array}\right),
      \end{equation}
  where $(\hat{\bf x}_{k,1}^T,\hat{\bf x}_{k,2}^T)^T=\hat{\bf x}_k^T=
  (x_k(n)W_{MN}^{kn})_{0\leq n\leq M}$ and 
  can be thought of as a modulated ${\bf x}_k$ for $0\leq k\leq N-1$. 

  From (\ref{10}), one can see that for any given information symbol vector
  ${\bf x}_{k,1}$, one can always find or  solve for $\hat{\bf x}_{k,2}$
  and thus a precoded ${\bf x}_{k,2}$ can be obtained
  by demodulating $W_{MN}^{kn}$ from $\hat{\bf x}_{k,2}$ such that
  the discrete frequency components in ${\bf y}_{k,2}$ of the
  discrete VOFDM signal ${\bf X}$  in (\ref{3})  are $0$.

  One remark is that 
  since the above precoding is done for a vector (or vector-wisely),
  when the vector size $M$ is not
    large, the precoding complexity is low.
  Another remark  is that in the above precoding,
  the number of variables in the precoded part ${\bf x}_{k,2}$
  is the same as that of the number of zeros in the spectrum vector 
  ${\bf y}_{k,2}$. In this case, the number of free information symbols
  in the unprecoded part ${\bf x}_{k,1}$ is the maximum. 
  However, as we shall see in the simulations in the next
  section, the precoded part ${\bf x}_{k,2}$ solved from the above
  linear system (\ref{10}) may have a higher power, which may cause
  a higher peak-to-average power ratio (PAPR) in the VOFDM signal.
  In order to have more degrees of freedom in designing the precoded
  ${\bf x}_{k,2}$ and thus to have a smaller PAPR, less 
  uncoded information symbols in ${\bf x}_{k,1}$, i.e., a
  smaller size vector ${\bf x}_{k,1}$, can be chosen and thus,
  more  variables in the precoded part ${\bf x}_{k,2}$, i.e.,
  a larger size vector ${\bf x}_{k,2}$, can
  be used/allocated to achieve the pre-set zero spectrum
  vector ${\bf y}_{k,2}$, i.e., to have ${\bf y}_{k,2}=0$. 

  \section{Simulations}\label{sec3}

  In this section we present some simple simulations to verify the theoretical
  results we have obtained above.

\begin{figure}%[t!]
\centering
\includegraphics[scale=0.53]{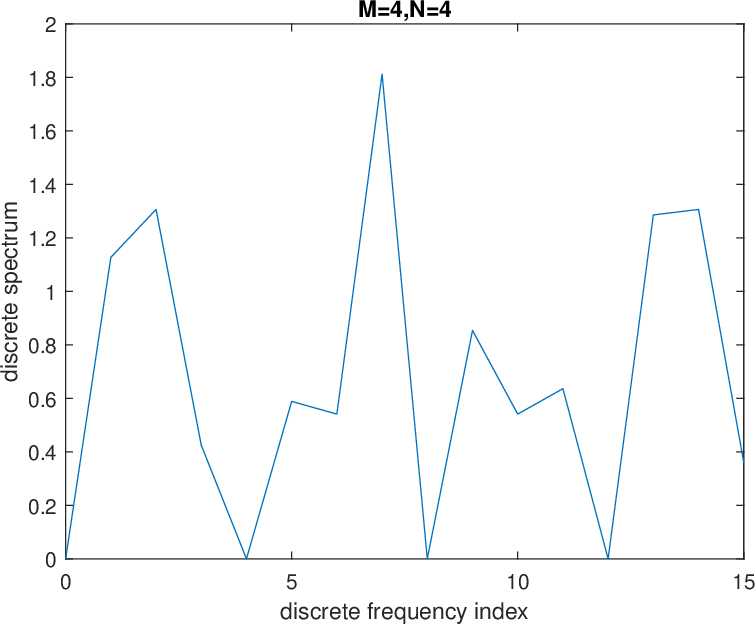}
\caption{Spectrum of a VOFDM signal of length $16$ with $M=N=4$ and the first
  information symbol vector ${\bf x}_0$ is set to $0$.}
\label{fig1}
\end{figure}

  For the illustratuon purpose, we first consider the case when $M=N=4$, i.e.,
 the vector size $M=4$ and the IFFT size $N=4$. In this case we set the first
  information symbol vector ${\bf x}_0$ to $0$, then according to Theorem 1,
  the frequency component vector ${\bf y}_0$ is $0$, whose $4$ components
  are evenly distributed over the total $16$ frequency points,
  which can be seen from Fig. \ref{fig1}. 
 
\begin{figure}%[t!]
\centering
\includegraphics[scale=0.53]{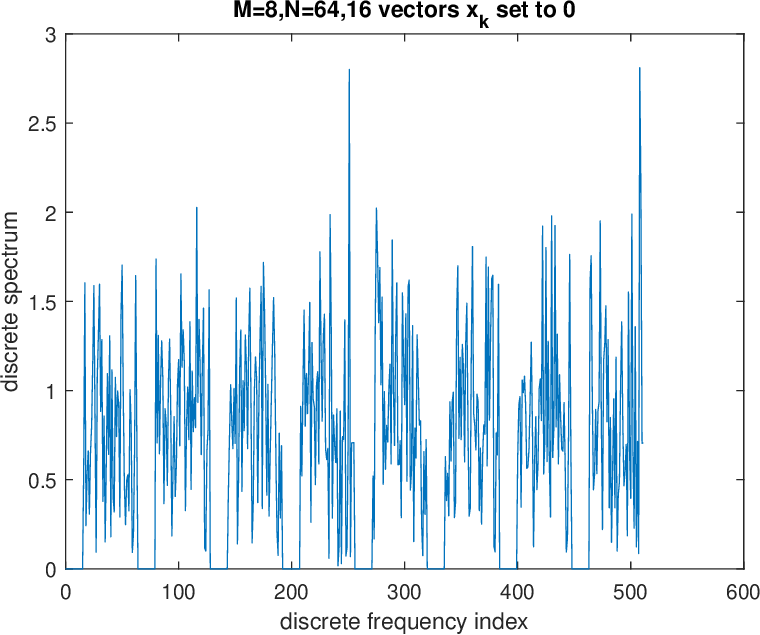}
\caption{Spectrum of a VOFDM signal of length $512$ with $M=8$ and $N=64$,
  and the first $16$ information symbol vectors ${\bf x}_k$, $0\leq k\leq 15$,
  are set to $0$.}
\label{fig2}
\end{figure}

We next consider the case when $M=8$ and $N=64$, i.e., the vector size is $M=8$
and the IFFT size is $N=64$. In this case we set the first $16$ consecutive
information symbol 
vectors ${\bf x}_k$ for $0\leq k\leq 15$ all $0$. According to Theorem 1,
the first $16$ consecutive frequency component vectors ${\bf y}_k$ for $0\leq k\leq 15$ are all 0, which can be seen from Fig. \ref{fig2}.

\begin{figure}%[t!]
\centering
\includegraphics[scale=0.53]{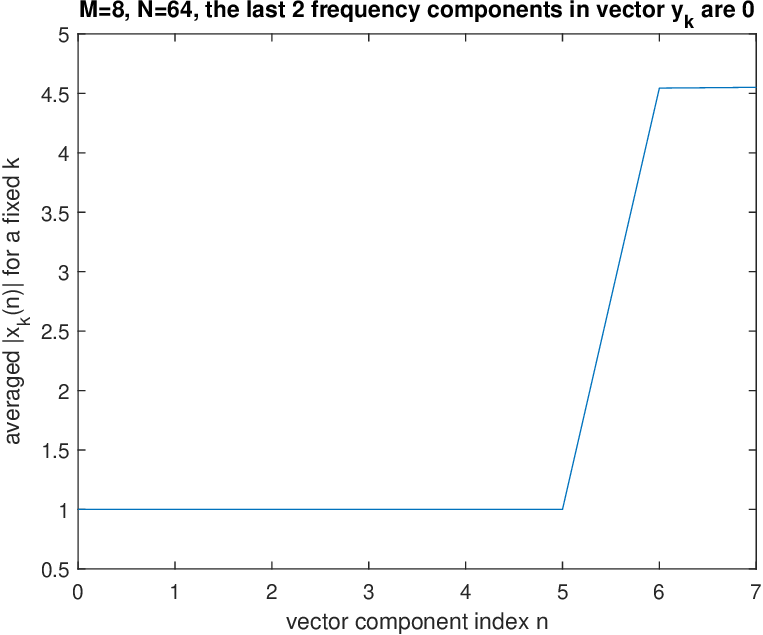}
\caption{Averaged magnitudes  of a precoded signal ${\bf x}_k$ with $M=8$ and $N=64$, where the last $2$ frequency components in ${\bf y}_k$
  are set to $0$.}
\label{fig3}
\end{figure}

\begin{figure}%[t!]
\centering
\includegraphics[scale=0.53]{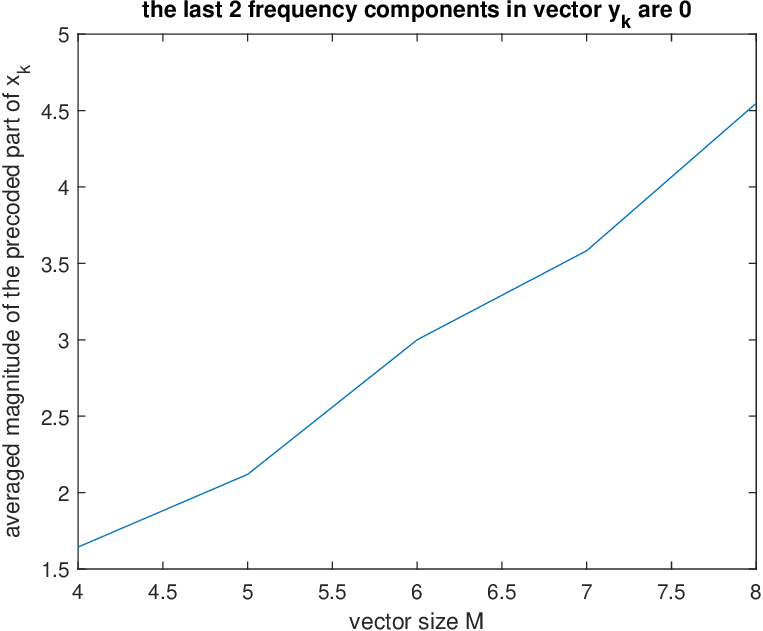}
\caption{Averaged magnitudes  of the precoded part ${\bf x}_{k,2}$ for
  different vector sizes $M$, where $N=64$ and the last $2$ frequency components in ${\bf y}_k$  are set to $0$.}
\label{fig33}
\end{figure}

Since the precoded part ${\bf x}_{k,2}$ for a given information symbol vector
${\bf x}_{k,1}$ and a zero setting in ${\bf y}_{k,2}$ in ${\bf y}_k$ is solved
via the linear system (\ref{10}), its power is a variable. We next set the last
two spectrum components  in ${\bf y}_k$ to $0$
for a fixed index $k$. We then component-wisely calculate the
averaged magnitudes of the signal vector ${\bf x}_k=({\bf x}_{k,1}^T, {\bf x}_{k,2}^T)^T$ including the random information symbol vector ${\bf x}_{k,1}$
with BPSK signals of amplitudes $\pm 1$ and the precoded 
    part ${\bf x}_{k,2}$ that includes the last components of ${\bf x}_k$.
The average is done over $1000$ random trials. The averaged magnitudes 
of the signal vector 
         ${\bf x}_k$ are shown in Fig. \ref{fig3} component-wisely.
The averaged magnitudes of the precoded part ${\bf x}_{k,2}$
for different vector sizes $M$ are plotted in Fig. \ref{fig33},
where only the last two components in vector ${\bf x}_k$ are precoded.

\begin{figure}%[t!]
\centering
\includegraphics[scale=0.53]{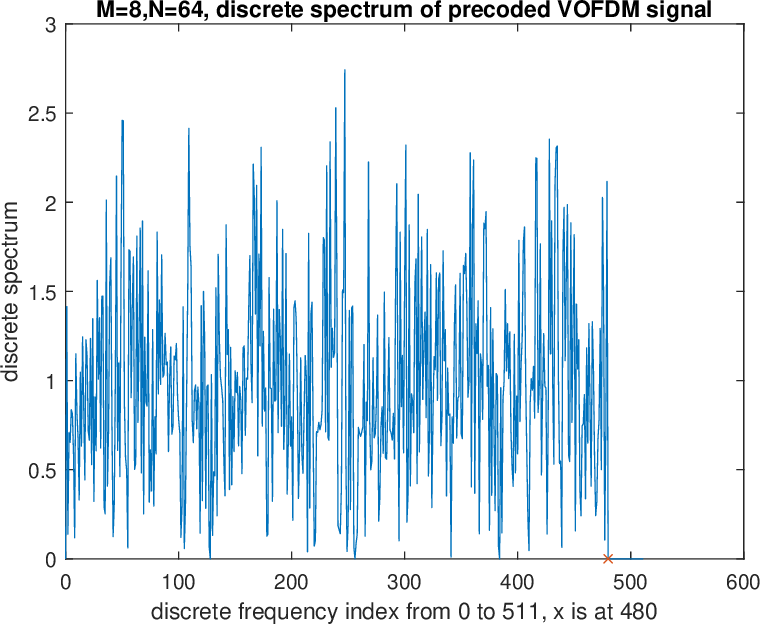}
\caption{Frequency spectrum  of a VOFDM signal with the last $32$
  frequency components $0$.}
\label{fig4}
\end{figure}

\begin{figure}%[t!]
\centering
\includegraphics[scale=0.53]{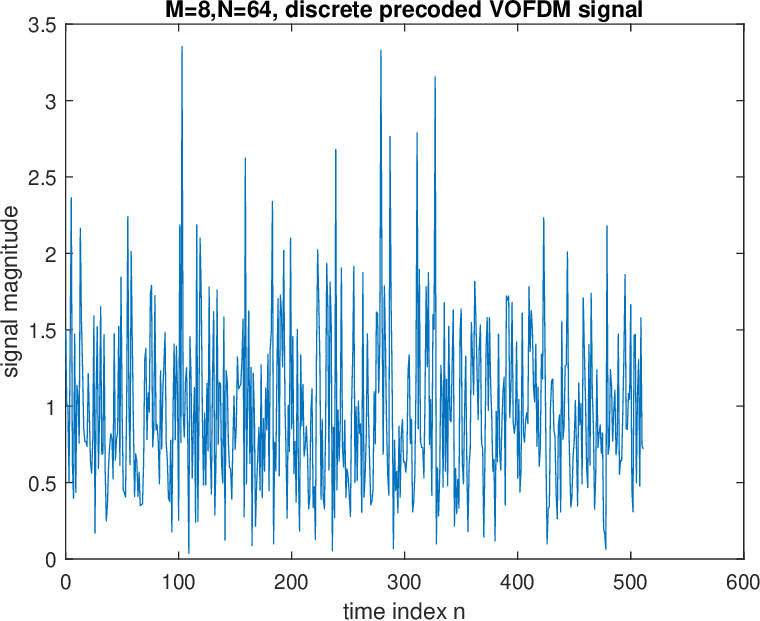}
\caption{The VOFDM signal with its spectrum shown in Fig. \ref{fig4}.}
\label{fig44}
\end{figure}

We next want to avoid the last $32$ frequencies in a discrete
spectrum of total $512$ frequency points for a VOFDM system
with vector size $M=8$ and IFFT size $N=64$.
This can be done by setting the last frequency components of the
second half  $32$ frequency response vectors ${\bf y}_k$, $32\leq k\leq 63$,
to $0$. After doing so, the frequency spectrum of such a VOFDM signal
is shown in Fig. \ref{fig4}, where one can clearly see that
the spectra at the last $32$ frequency points are $0$.
For this example, according to Theorem 1 only the last components
of the second half signal vectors
${\bf x}_k$, $32\leq k\leq 63$,
need to be precoded and the first $31$ components
of each of  these vectors are free.
In other words, for any of these $32$ information symbol vectors,
only the last component needs to be precoded. 
Also, the first $32$ information
symbol vectors ${\bf x}_k$, $0\leq k\leq 31$, are free as well.
Its corresponding discrete VOFDM signal magnitudes of length $512$ 
in time domain before CP are shown in Fig. \ref{fig44}, where
one can see that the PAPR may be high. 

\begin{figure}%[t!]
\centering
\includegraphics[scale=0.53]{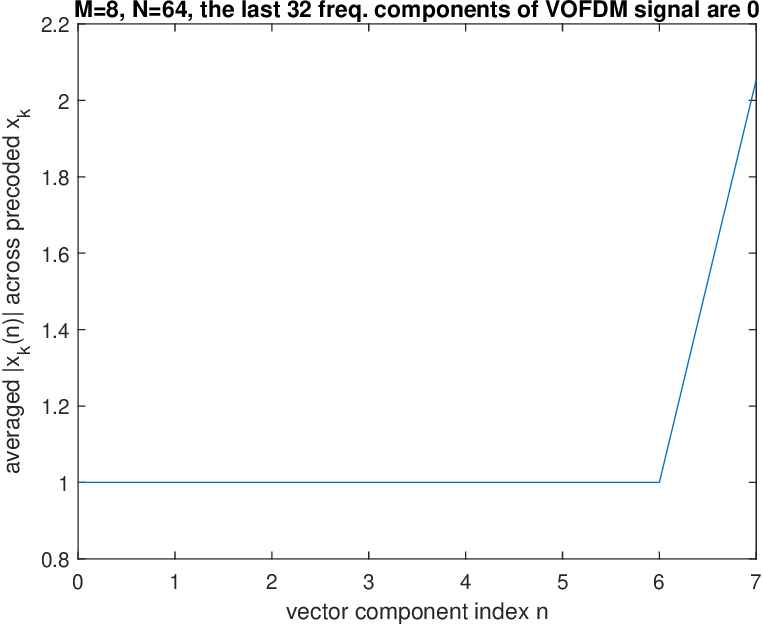}
\caption{Averaged magnitudes of  the last $32$ information
  symbol vectors ${\bf x}_k$, $31\leq k\leq 63$, where the last components
  of these vectors are precoded.}
\label{fig5}
\end{figure}

The averages
of the magnitudes 
across the second half vectors ${\bf x}_k$, $32\leq k\leq 63$, 
are shown in Fig. \ref{fig5} component-wisely, where the first $31$ components
of every vector ${\bf x}_k$, $32\leq k\leq 63$,  and  any of the first half vectors ${\bf x}_k$, $0\leq k\leq 31$, all take BPSK signals with amplitudes $\pm 1$
randomly. 

\section{Conclusion}\label{sec4}

In this paper, we have first obtained a linear relationship
between an information symbol vector and a corresponding
frequency response vector of the discrete frequency spectrum
of the VOFDM signal. It is the $M$-point DFT of the modulated
information symbol vector. From this linear relationship,
it is easy to see that a zero information symbol vector leads
to a zero frequency response vector of the spectrum of the OFDM
signal. Also,  information symbol vectors can be locally  precoded
so that any components in the frequency response vectors
of the spectrum of the VOFDM signals can be set to $0$.
Thus, any discrete spectrum components can be set to $0$ in a VOFDM
system similar to an OFDM system.
Since the precoding is done for a vector (or vector-wisely),
when the vector size
is not large, the complexity is low.
These results have
been verified by numerical simulations.

\end{document}